\documentclass[
aps,prd,
12pt,
nopreprintnumbers,
showpacs,
eqsecnum,
nofootinbib
]{revtex4-1}

\usepackage{graphicx}
\usepackage{amssymb}

\begin{document}

\title{A Newtonian Analysis of Multi-scalar Boson Stars with Large
Self-couplings
\\ {\small ---A Poor Person's Approach to Flat Galaxy Rotation Curves---}}
\author{Nahomi Kan}\email[]{kan@gifu-nct.ac.jp}
\affiliation{National Institute of Technology, Gifu College,
Motosu-shi, Gifu 501-0495, Japan
}
\author{Kiyoshi Shiraishi}\email[]{shiraish@yamaguchi-u.ac.jp}
\affiliation{
Graduate School of Sciences and Technology for Innovation, Yamaguchi
University, Yamaguchi-shi, Yamaguchi 753--8512, Japan}
\date{\today}

\begin{abstract}
We study solutions for boson stars in the multi-scalar field theory with
global symmetry $[U(1)]^N$.  The properties of the boson stars are
investigated by the Newtonian approximation with the
large coupling limit.
Our purpose is to study the models bringing about exotic mass
distributions which explain flat rotation curves of galaxies.
We propose plausible models in which coupling matrices are associated with
various graphs in graph theory.
\end{abstract}



\maketitle

\section{Introduction}
\label{sec1}

From the perspective of particle cosmology, the boson star
\cite{Jetzer,LM,SM,LP} is one of the
 candidates for dark matter \cite{R1SL1,R1SL2,LK,TCL,ST,ABBR,MA,BBAP,FG,Sin,JS}. 
The ($1+3$)-dimensional boson
star was studied as the simplest object of a self-gravitating system and
the Newtonian treatment of gravitating bosons has been often discussed
(for example, see Refs.~\cite{Jetzer,SM,LP,LK,MA,FG,Sin,JS,KSrecent}).  

Boson stars or boson halos have been studied sometimes
in expectation of  solving 
the flat rotation curve  problem \cite{LK,TCL,ST,ABBR,MA,BBAP,FG,Sin,JS}.%
\footnote{The flat rotation curve problem was also approched by Schunck
\cite{Schunck}, who considered a configuration of a
massless scalar field with infinite range.}
Some authors have attempted to explain rotation curves of galaxies
by assuming the existence of a galactic-scale boson star located at the center of
a galaxy. In order to fit the observed data,
the mass density of the boson star needs to be widely distributed.
Such configurations can be constructed by the models such as 
boson stars with a scalar field in an excited state or rotating boson stars
and show
better agreement with the galaxy rotation curves,  
but these boson stars are sometimes
unstable. Whereas
Newtonian boson stars in only one ground state are stable,
it is difficult to illustrate the realistic rotation curve owing to their
compact density distribution. Alternative models to solve these problems have been
investigated by Matos and Ure\~na-L\'opez~\cite{MA},  and by Bernal {\it et
al.}~\cite{BBAP}. They considered the multi-state boson stars, {i.e.} a scalar
field both in ground and in excited states, with no (quartic) self-couplings.

In the present paper,
we consider scalar boson stars made of multi-species or charges, not multi-states
of a single scalar field.
We suggest the models which
contain multiple scalar fields with self- and mutual-couplings.%
\footnote{Interacting boson stars and Q-balls have been 
studied by Brihaye {\it et al.}~\cite{Brihaye1,Brihaye2,Brihaye3}
in the other context.}
It is shown that the configuration induced from the multi-bosons can improve the
flatness of the galactic rotation curve at the large scale.

We also claim that the coupling matrix can be built up by the knowledge of graph
theory. The model we propose contains $N$ scalar fields interacting with
oneself and with `adjacent' scalar fields on a graph. 
A similar type of many $U(1)$ interacting fields has been motivated by the
graph-oriented model with supersymmetry \cite{KKS}. 

To discuss the qualitative behavior of models such as a static and 
spherical boson star, we study the system of scalar fields
with large self-couplings \cite{colpi} in this paper. 
An understanding of the
basic aspects of general multi-boson systems is yet expected from this
approximation.

The present paper is organized as follows.
In Sec.~\ref{sNL}, we propose the action, the Hamiltonian, and the field
equations for a model of self-interacting, gravitating scalar fields in
the Newtonian limit. 
The large coupling limit of the model is defined in Sec.~\ref{sec3}.
In Sec.~\ref{sec4},  as the simplest case, analytic solutions for boson stars in
a bi-scalar theory are obtained, and relations among their physical
quantities is derived.
In Sec.~\ref{sec5}, we study boson stars in a multi-scalar theory
with only self-couplings of individual scalar fields, i.e., with a diagonal
coupling matrix. We propose symmetric models based on the graph theory,
which are compatible to deriving the desired profile of a multi-scalar boson star,
in Sec.~\ref{sec6}. In this section, we mainly concentrate ourselves on a model
associated with a complete graph. Other possible models based on the other types of
graphs are also mentioned. The last section \ref{so} is devoted to a summary and
future prospects.

\section{The multi-scalar model and its Newtonian limit}
\label{sNL}
We consider a system of interacting, gravitating complex scalar bosons
$\phi_i$ $(i=1,\dots, N)$
of equal mass $m$,%
\footnote{In the large coupling limit being treated in the next section, a possible
mass spectrum is absorbed into redefinition of scalar fields by construction.} 
governed by the following relativistic action (where $\hbar=c=1$); 
\begin{eqnarray}
S&=&\int d^4x\,{\sqrt{-g}}\,{\cal L}\nonumber \\
&=&\int
d^4x{\sqrt{-g}}\left[\frac{1}{16\pi G}R-\sum_{i=1}^N
\{|\partial\phi_i|^2+m^2|\phi_i|^2\}-
\frac{1}{2}\sum_{i=1}^N\sum_{j=1}^N\tilde{\lambda}_{ij}|\phi_i|^2
|\phi_j|^2\right]\,,
\end{eqnarray}
where $d^4x=dt\,d^3\mbox{\boldmath $r$}$ , $G$ is the Newton constant,
$R$ is the Ricci scalar, $g$ is the determinant of the metric $g_{\mu\nu}$
$(\mu,\nu=0,1,2,3)$,
$\tilde{\lambda}_{ij}=\tilde{\lambda}_{ji}$ $(i,j=1,\dots,N)$ are the dimensionless
real (self and mutual) coupling constants, and
$|\partial\phi_i|^2\equiv
g^{\mu\nu}(\partial_\mu\phi_i)^*(\partial_\nu\phi_i)$.
The action has global $[U(1)]^N$ symmetry.

By the variational principle, we derive the Einstein equation from the
action as
\begin{equation}
R^\mu_\nu-\frac{1}{2}\delta^\mu_\nu R=8\pi G T^\mu_\nu\,,
\end{equation}
where the energy-momentum tensor $T_{\mu\nu}$ in the system is given by
\begin{equation}
T_{\mu\nu}=\sum_i\Big[\partial_\mu\phi_i^*\partial_\nu\phi_i+
\partial_\nu\phi_i^*\partial_\mu\phi_i-
g_{\mu\nu}|\partial\phi_i|^2-g_{\mu\nu}
m^2|\phi_i|^2\Big]
-g_{\mu\nu}V(|\phi_i|^2)
\,,
\end{equation}
where the quartic interaction is denoted as
\begin{equation}
V(|\phi_i|^2)\equiv\frac{1}{2}\sum_{i,j}\tilde{\lambda}_{ij}|\phi_i|^2|\phi_j|^2\,.
\end{equation}
On the other hand, the equation of motion for each scalar field $\phi_i$ is given
by
\begin{equation}
\Box\phi_i-m^2\phi_i-\sum_{j}\tilde{\lambda}_{ij}|\phi_j|^2\phi_i=0\,,
\label{basiceq}
\end{equation}
where $\Box\phi\equiv\frac{1}{\sqrt{-g}}\partial_\mu
(\sqrt{-g}g^{\mu\nu}\partial_\nu
\phi)$ is the covariant d'Alembertian.

Throughout the present paper, we consider the gravitating system in the Newtonian
approximation. The Newtonian limit can be attained by assuming that the spacetime
metric in the weak field approximation can be written as
\begin{equation}
g_{00}\approx -\left(1+
{2\Phi}
\right)\,,\quad g_{ij}\approx \delta_{ij}\,,\quad 
g_{0i}=g_{i0}\approx 0\,,\quad\sqrt{-g}\approx
1\,,
\end{equation}
where $\Phi$ is the Newtonian gravitational potential.

Assuming further that the complex scalar field has a nearly harmonic time
dependence expressed by 
\begin{equation}
\phi_i
\approx\frac{1}{\sqrt{2m}}\psi_i(\mbox{\boldmath $r$}, t)\,e^{-imt}\,,
\end{equation}
we obtain the (non-linear) Schr\"odinger equation
\begin{equation}
i\dot{\psi}_i=-\frac{1}{2m}\nabla^2\psi_i+m\Phi\psi_i+
\frac{1}{4}\sum_j\frac{\tilde{\lambda}_{ij}}{m^2}|\psi_j|^2\psi_i\,
\end{equation}
as the Newtonian limit of Eq.~(\ref{basiceq}), where $\nabla^2$ is the
Laplacian in the flat space and the dot ($\dot{~}$) indicates the time derivative.
In the present limit, the Einstein equations reduce to the Poisson
equation
\begin{equation}
\nabla^2\Phi=4\pi Gm\sum_i|\psi_i|^2\,.
\end{equation}

The Newtonian treatment of the Lagrangian and Hamiltonian is as follows.
We find the following Newtonian action in the limit:
\begin{equation}
S\cong\int dt\, d^3\mbox{\boldmath $r$}
\left[-\frac{1}{8\pi G}({\nabla}
\Phi)^2+\sum_i\left\{i\psi_i^*\dot{\psi}_i-
\frac{1}{2m}|\nabla\psi_i|^2-m\Phi |\psi_i|^2\right\}-
\frac{1}{8}\sum_{i,j}\frac{\tilde{\lambda}_{ij}}{m^2}|\psi_i|^2|\psi_j|^2
\right]\,,
\end{equation}
where $(\nabla\Phi)^2\equiv {\bf\nabla}{\Phi}\cdot{\bf\nabla}{\Phi}$
and the symbol $\cong$ indicates that some surface terms have been
omitted.
Therefore, the Hamiltonian of the system is derived as
\begin{eqnarray}
& &\hat{H}(\psi_i,\Phi)=\int d^3\mbox{\boldmath $r$}\,{\cal H}\nonumber \\
&=&\int
d^3\mbox{\boldmath
$r$}
\left[\frac{1}{8\pi G}({\nabla}
\Phi)^2+\sum_i\left\{
\frac{1}{2m}|\nabla\psi_i|^2+m\Phi |\psi_i|^2\right\}+
\frac{1}{8}\sum_{i,j}\frac{\tilde{\lambda}_{ij}}{m^2}|\psi_i|^2|\psi_j|^2
\right]\,.
\label{NE}
\end{eqnarray}

The number of particles of the $i$-th species
is expressed as
\begin{equation}
\hat{n}_i\equiv\int d^3\mbox{\boldmath $r$}\,|\psi_i|^2\,.
\label{ndef}
\end{equation}

In addition, we require the condition $\hat{n}_i=n_i$,
i.e., the condition that the system contains $n_i$ scalar bosons of the $i$-th
species. Then, we consider
${\delta}\{\hat{H}-\sum_i\mu_i(\hat{n}_i-n_i)\}=0$ as equations
for the scalar fields in the mean field approximation, where $\mu_i$
are Lagrange multipliers and can be interpreted as chemical potentials for 
corresponding bosons.

Now, we obtain coupled equations for the stationary gravitational
field and the scalar fields as follows:
\begin{eqnarray}
& &\nabla^2\Phi=4\pi G m\sum_i|\psi_i|^2\,,
\label{ps1}
 \\
& &-\frac{1}{2m}\nabla^2\psi_i+m\Phi\psi_i+
\frac{1}{4}\sum_j\frac{\tilde{\lambda}_{ij}}{m^2}|\psi_j|^2
\psi_i=\mu_i\psi_i\,.\label{ps2}
\end{eqnarray}
Therefore, the system is reduced in the Newtonian limit to the
(non-linear) Schr\"odinger--Poisson system.

It is notable that the field equations (\ref{ps1}) and (\ref{ps2}) 
are invariant under a common shift of potentials
\begin{equation}
\Phi\rightarrow\Phi+\Phi_0\,,\quad \mu_i\rightarrow \mu_i+m\Phi_0\,.
\label{shiftinv}
\end{equation}
Therefore, we can choose $\Phi=0$ at the spatial infinity for a compact boson star,
even after solving the field equations.

In the subsequent sections of this paper, we will restrict ourselves on
the large coupling limit for compact objects, which will be defined in the next
section.

\section{The large coupling limit and the field equations}
\label{sec3}
Here, we consider the large coupling limit \cite{colpi}.
It is incidentally known that the large coupling leads to a large scale boson star.
First we define the matrix of couplings $\Lambda_{ij}$ as
\begin{equation}
\Lambda_{ij}\equiv\frac{1}{8\pi Gm^2}\tilde{\lambda}_{ij}\,.
\end{equation}
In addition, we introduce the following quantities
\begin{equation}
\mbox{\boldmath $r$}_*\equiv\frac{m}{\sqrt{\Lambda}}\mbox{\boldmath
$r$}\,,\quad
\Psi_i\equiv\sqrt{\frac{4\pi G\Lambda}{m}}\psi_i\,,\quad
{u_i}\equiv\frac{\mu_i}{m}\,,
\label{scale}
\end{equation}
where $\Lambda$ is a typical scale of $\Lambda_{ij}$.
Then, the set of equations reduces to the simple form
\begin{eqnarray}
& &\nabla^2_*\Phi=\sum_i|\Psi_i|^2\,,\label{Poi}
\label{33}
 \\
& &-\frac{1}{2\Lambda}\nabla^2_*\Psi_i+\Phi\Psi_i+
\frac{1}{2}\sum_jC_{ij}|\Psi_j|^2
\Psi_i=u_i\Psi_i\,,
\label{LCL}
\end{eqnarray}
where $C_{ij}\equiv\frac{1}{\Lambda}\Lambda_{ij}$.
$\nabla^2_*$ is the rescaled Laplacian expressed in terms of the
coordinate
$\mbox{\boldmath
$r$}_*$.

In the limit of $\Lambda\rightarrow\infty$, equation (\ref{LCL})
further reduces to the algebraic equation:
\begin{equation}
\Phi\Psi_i+
\frac{1}{2}\sum_jC_{ij}|\Psi_j|^2
\Psi_i=u_i\Psi_i\,.
\label{LL}
\end{equation}
In this paper, we only consider the case with the large coupling limit.
It is interesting to note that the equation (\ref{LL}) as well as the Poisson
equation (\ref{33}) are invariant under the following scale transformation:
\begin{equation}
\Phi\rightarrow\lambda^2\Phi\,,\quad u_i\rightarrow\lambda^2 u_i\,,
\quad \Psi_i\rightarrow\lambda\Psi_i\,,
\label{scaletransf}
\end{equation}
where $\lambda$ is a constant.

We further define normalized (particle number) density functions as
\begin{equation}
\rho_i\equiv|\Psi_i|^2\,.
\label{37}
\end{equation}
Then, the particle number of the $i$-th scalar boson is given by
\begin{equation}
n_i=\frac{\sqrt{\Lambda}}{4\pi Gm^2}\int d^3\mbox{\boldmath $r$}_*\,
\rho_i(\mbox{\boldmath
$r$}_*)\,.
\label{np}
\end{equation}

The Newtonian energy $E$ of the system in the
large coupling limit can be expressed from Eq.~(\ref{NE}) as
\begin{equation}
E=\frac{\sqrt{\Lambda}}{4\pi Gm}\int d^3\mbox{\boldmath $r$}_*
\left[\frac{1}{2}({\nabla}_*
\Phi)^2
+\Phi \sum_i|\Psi_i|^2+
\frac{1}{4}\sum_{ij}C_{ij}|\Psi_i|^2|\Psi_j|^2
\right]\,.
\label{energy}
\end{equation}
Substituting the solution of Eqs.~(\ref{33}), (\ref{LL}) and
(\ref{37}) into this equation, we obtain
\begin{eqnarray}
E&\cong&\frac{\sqrt{\Lambda}}{4\pi Gm}\int d^3\mbox{\boldmath $r$}_*
\left[-\frac{1}{2}\Phi\nabla_*^2\Phi+\Phi\sum_i\rho_i+
\frac{1}{4}\sum_{ij}C_{ij}\rho_i\rho_j
\right]\nonumber \\
&=&\frac{\sqrt{\Lambda}}{4\pi Gm}\int d^3\mbox{\boldmath $r$}_*
\left[\frac{1}{2}\Phi\sum_i \rho_i+
\frac{1}{4}\sum_{ij}C_{ij}\rho_i\rho_j
\right]
\nonumber \\
&=&\frac{\sqrt{\Lambda}}{4\pi Gm}\int d^3\mbox{\boldmath $r$}_*
\left[\frac{1}{2}\sum_iu_i
\rho_i\right]=\frac{1}{2}m\sum_in_i u_i=\frac{1}{2}\sum_in_i\mu_i\,.
\label{sphe}
\end{eqnarray}
For a compact object, the energy becomes negative ($E<0$).

The mass of the boson star is given in the present Newtonian scheme by
\begin{equation}
M=m\sum_in_i+E=\sum_in_i\left(m+\frac{1}{2}\mu_i\right)
=m\sum_in_i\left(1+\frac{1}{2}u_i\right)\,.
\end{equation}

Finally in the present section, we consider the field equations for static,
spherically symmetric solutions of the system.  Then, the field equations
(\ref{33}) and (\ref{LCL}) can be rewritten as
\begin{equation}
\frac{1}{x}\frac{d^2}{dx^2}(x\Phi(x))=S(x)\,,
\end{equation}
\begin{equation}
\left(\Phi(x)+
\frac{1}{2}\sum_jC_{ij}\rho_j(x)
-u_i\right)\rho_i(x)=0\,.
\end{equation}
where
\begin{equation}
S(x)\equiv\sum_{i}\rho_i(x)\,,\quad
x\equiv\sqrt{\mbox{\boldmath $r$}_*\cdot\mbox{\boldmath $r$}_*}\,.
\end{equation}

\section{Non-Relativistic spherical bi-scalar boson stars 
in large coupling limit}
\label{sec4}

We first examine the simplest case, static and spherical solutions for boson stars
in the system with two scalar fields.
There are not single, but two scalar conserved charges for boson 1 and boson 2.

We consider a boson star model, in which two scalar fields
with the $2\times 2$ coupling matrix  $C_{ij}$.
The field equations for a spherical configuration derived in the previous section
become as follows:
\begin{eqnarray}
&&~~~
\frac{1}{x}\frac{d^2}{dx^2}(x\Phi(x)) = \rho_1(x)+\rho_2(x)
\label{poi}\,,\\
&&\left[\Phi(x) + \frac{1}{2} \left(
{C_{11}}\rho_1(x)+{C_{12}}\rho_2(x)\right)-u_1\right]\rho_1(x)=0\,,
\label{ae1}\\
&&\left[\Phi(x) + \frac{1}{2} \left(
{C_{21}}\rho_1(x)+{C_{22}}\rho_2(x)\right)-u_2\right]\rho_2(x)=0
\label{ae2}\,,
\end{eqnarray}
where we should remember that $C_{12}=C_{21}$.
It is assumed that  $C_{11}C_{22}-C_{12}^2>0$, $C_{11}>0$,
for the positive definite $V(|\phi_i|^2)$.

In the large coupling limit, the exponential asymptotic distribution of the
scalar density is suppressed \cite{colpi}.
Thus, for the multi-scalar case, it is notable that there are regions where
densities of some species of scalar fields vanish.
Assuming the normalized densities $\rho_1 \neq 0$ for the boson
1 and
$\rho_2
\neq 0$ for the boson 2 in the core region of the boson star (in the vicinity of
the coordinate origin), the algebraic equations (\ref{ae1}) and (\ref{ae2}) become
\begin{equation}
\frac{1}{2}
\left(
\begin{array}{cc}
{C_{11}} & {C_{12}}\\
{C_{12}} & {C_{22}}
\end{array}\right)\left(
\begin{array}{c}
\rho_1(x)\\
\rho_2(x)
\end{array}\right)=
\left(
\begin{array}{c}
-\Phi(x)+{u_1}\\
-\Phi(x)+{u_2}
\end{array}\right)\,,
\end{equation}
and quite equivalently
\begin{equation}
\left(
\begin{array}{c}
\rho_1(x)\\
\rho_2(x)
\end{array}\right)=\frac{2}{C_{11}C_{22}-{C_{12}}^2}
\left(
\begin{array}{cc}
{C_{22}} & -{C_{12}}\\
-{C_{12}} & {C_{11}}
\end{array}\right)\left(
\begin{array}{c}
-\Phi(x)+{u_1}\\
-\Phi(x)+{u_2}
\end{array}\right)\,.
\end{equation}
Thus, we define the total `density' $S_2$ and it can be expressed as
\begin{equation}
S_2(x)\equiv\rho_1(x)+\rho_2(x)=-2\frac{ C_{11} - 2 C_{12}+ C_{22}}{C_{11}
C_{22}-{C_{12}}^2}\Phi(x)+2\frac{(C_{22} -  C_{12})u_1+ (C_{11}-C_{12})u_2}{C_{11}
C_{22}-{C_{12}}^2}\,.
\end{equation}
Using the Poisson equation (\ref{poi}), we obtain the differential equation on
$S_2$ as
\begin{equation}
\frac{1}{x}\frac{d^2}{dx^2} (xS_2(x))=-\omega_2^2S_2(x)\,,
\end{equation}
where the positive constant $\omega_2$ satisfies
\begin{equation}
\omega_2^2 = 2 
\frac{ C_{11} - 2 C_{12}+ C_{22}}{C_{11}
C_{22}-{C_{12}}^2}.
\end{equation}

The solution of the above equation takes the form
\begin{equation}
S_2(x)=A_2\frac{\sin \omega_2 x}{x}\,, 
\end{equation}
because the center of the boson star at $x=0$ should be nonsingular.
Here, the scale factor $A_2$ is a positive constant.
Then, the gravitational potential is written as
\begin{equation}
\Phi(x) +  const. = -A_2\frac{\sin \omega_2 x}{\omega_2^2 x}\,,
\end{equation}
in this core region.

In the outer region of the star, where $\rho_1\ne 0$ and $\rho_2= 0$,
field equations become
\begin{equation}
\Phi(x)+\frac{1}{2}
{C_{11}}\rho_1(x)=u_1\,,
\end{equation}
and taking account of the Poisson equation (\ref{poi}), we find
\begin{equation}
\frac{1}{x}\frac{d^2}{dx^2} (x\rho_1(x))=-\omega_1^2\rho_1(x)\,,
\end{equation}
where the positive constant $\omega_1$ satisfies
\begin{equation}
\omega_1^2 =  
\frac{2}{C_{11}}.
\end{equation}
Then the solution can be written as
\begin{equation}
\rho_1(x)=A_1 \frac{\sin (\omega_1 x + \delta_1)}{x}\equiv S_1(x)\,,
\end{equation}
where $\delta_1$ and $A_1$ are constants.
The gravitational potential then becomes
\begin{equation}
\Phi(x) +  const. = -A_1\frac{\sin (\omega_1 x + \delta_1)}{\omega_1^2x}\,,
\end{equation}
in this outer region.

We define the boundary of two regions, where $\rho_2\neq 0$ and $\rho_2=0$, as
$x=x_2$ and define the outermost surface of the boson star as $x=x_1$, where
$\rho_1=\rho_2=0$. Then, we find that $\rho_1\neq 0$ and $\rho_2\neq 0$ for $0\le
x\le x_2$,
 $\rho_1\neq 0$ and $\rho_2=0$ for $x_2\le x\le x_1$, and 
 $\rho_1=\rho_2=0$ for $x\ge x_1$.

Thus,
at $x=x_2$, since the total density is continuous,
\begin{equation}
A_2 {\sin \omega_2 x_2}=A_1 {\sin (\omega_1 x_2 + \delta_1)}\,,
\label{e1}
\end{equation}
is satisfied.%
\footnote{Note that there is no condition on the first derivative of $\rho_1(x)$
and
$\rho_2(x)$ at $x=x_2$.}
Further, since the gravitational force which is derived from the derivative of the
potential $\Phi'(x)\equiv\frac{d\Phi}{dx}$ varies continuously even at the
boundary, the equality
\begin{equation}
A_2 \frac{\omega_2x_2\cos \omega_2 x_2-\sin\omega_2 x_2}{\omega_2^2x_2^2}=A_1
\frac{\omega_1x_2\cos (\omega_1 x_2 +
\delta_1)-\sin(\omega_1 x_2 +
\delta_1)}{\omega_1^2x_2^2}\,,
\label{e2}
\end{equation}
should be hold.
Combining two equalities (\ref{e1}) and  (\ref{e2}), we obtain
\begin{equation}
\frac{\omega_2x_2\cot \omega_2 x_2-1}{\omega_2^2}=
\frac{\omega_1x_2\cot (\omega_1 x_2 +
\delta_1)-1}{\omega_1^2}\,,
\label{deltas}
\end{equation}
and this equation tells us the value for $\delta_1$ if $x_2$ is given.
On the other hand, the outer radius of the boson star $x_1$ is determined by the
simple equation
\begin{equation}
\sin(\omega_1x_1+\delta_1)=0\,.
\end{equation}

At last, the analytic solution can be obtained, for a given $x_2$, as follows.
\begin{equation}
\rho_1(x)=\left\{
\begin{array}{cc}
\frac{S(0)}{C_{11}-2C_{12}+C_{22}}\left[\frac{(C_{22}-C_{12})\sin\omega_2
x}{\omega_2x} +\frac{(C_{11}-C_{12})\sin\omega_2 x_2}{\omega_2x_2}
\right] & (0\le x\le x_2)\\
\frac{S(0)\sin\omega_2 x_2}{\sin(\omega_1 x_2+\delta_1)}\frac{\sin(\omega_1
x+\delta_1)}{\omega_2x} & (x_2\le x\le x_1)\\
0 & (x\ge x_1)
\end{array}
\right.\,,
\end{equation}
\begin{equation}
\rho_2(x)=\left\{
\begin{array}{cc}
\frac{S(0)(C_{11}-C_{12})}{C_{11}-2C_{12}+C_{22}}\left[\frac{\sin\omega_2
x}{\omega_2x} -\frac{\sin\omega_2 x_2}{\omega_2x_2}
\right] & (0\le x\le x_2)\\
0 & (x\ge x_2)
\end{array}
\right.\,,
\end{equation}
where $\delta_1$ is considered as the solution of Eq.~(\ref{deltas}), also
hereafter in the present section. Note that $S(0)=\rho_1(0)+\rho_2(0)$.

The Newtonian gravitational potential is solved as
\begin{equation}
\frac{1}{S(0)}\Phi(x)=\left\{
\begin{array}{cc}
-\frac{\sin\omega_2 x}{\omega_2^3x}-\frac{\sin\omega_2
x_2}{\omega_1\omega_2\sin(\omega_1
x_2+\delta_1)}-\left(\frac{1}{\omega_1^2}-\frac{1}{\omega_2^2}\right)\frac{\sin\omega_2x_2}{\omega_2x_2}
& (0\le x\le x_2)\\ -\frac{\sin\omega_2 x_2}{\omega_1\omega_2\sin(\omega_1
x_2+\delta_1)}\left[\frac{\sin(\omega_1 x+\delta_1)}{\omega_1x}+1\right] & (x_2\le
x\le x_1)\\ -\frac{\sin\omega_2 x_2}{\omega_1\omega_2\sin(\omega_1
x_2+\delta_1)}\frac{x_1}{x} & (x\ge x_1)
\end{array}
\right.\,,
\end{equation}
where we set $\Phi(\infty)=0$ by using the shift invariance (\ref{shiftinv}) for
potentials. 

The chemical potentials can also be obtained as
\begin{eqnarray}
\frac{1}{S(0)}u_1&=&-\frac{\sin\omega_2 x_2}{\omega_1\omega_2\sin(\omega_1
x_2+\delta_1)}\,,\\
\frac{1}{S(0)}u_2&=&-\frac{\sin\omega_2 x_2}{\omega_1\omega_2\sin(\omega_1
x_2+\delta_1)}-\frac{C_{11}-C_{12}}{2}\frac{\sin\omega_2 x_2}{\omega_2x_2}\,,
\end{eqnarray}
which are is always negative, as for a bound state.

\begin{figure}[ht]
\centering
\includegraphics[keepaspectratio=true,width=7.5cm
]{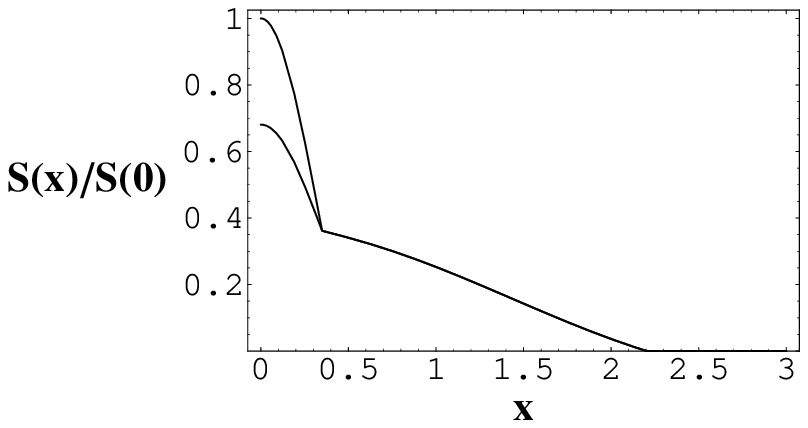}
\caption{
The behavior of the density profile of scalar bosons
as the function of the rescaled distance
$x$.
The upper curve 
represents $S_2=\rho_1+\rho_2$ and the lower curve appeared for $x<x_2=0.35$
represents $\rho_1$. 
The couplings are set as $C_{11}=C_{22}=1$,
$C_{12}=0.9$. The kink in the curve is due to the approximation of the large
coupling limit.}
\label{fig01}
\centering
\includegraphics[keepaspectratio=true,width=7.5cm
]{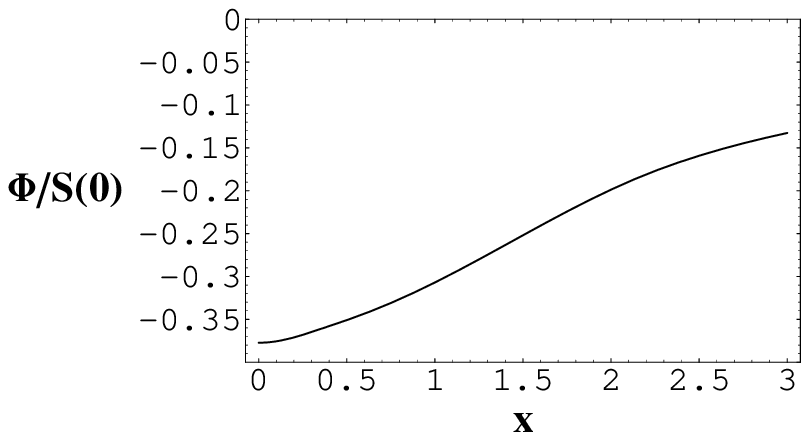}
\caption{
The behavior of the gravitational potential $\Phi(x)$
as the function of the rescaled distance
$x$.
The
couplings are set as $C_{11}=C_{22}=1$,
$C_{12}=0.9$ and $x_2$ is chosen as $x_2=0.35$. 
}
\label{fig02}
\end{figure}

A typical profile of the bi-scalar boson star and the gravitational potential are
exhibited in FIG.~\ref{fig01} and FIG.~\ref{fig02}, respectively, where the
couplings are set as $C_{11}=C_{22}=1$,
$C_{12}=0.9$ and $x_2$ is chosen as $x_2=0.35$. 
In FIG.~\ref{fig01}, one can find that the density profile has a kink structure at
$x=x_2$. This is due to the approximation of the large coupling limit, since 
the approximation is equivalent to assuming strong but very short-range repulsive
forces among bosonic particles. As we will see later, density profiles seem
to be almost smooth in multi-scalar cases.

Due to the scale invariance, we obtain a general solution by
multiplying an arbitrary common constant with the above set of the solution
for $\rho_1(x)$, $\rho_2(x)$, $u_1$, $u_2$ and $\Phi(x)$.

The fraction of the particle number of the boson 2, which lives inner region of
the boson star, is expressed as
\begin{equation}
\frac{n_2}{n}=\frac{(C_{11}-C_{12})\omega_1
\sin(\omega_1x_2+\delta_1)}{C_{11}-2C_{12}+C_{22}}
\frac{3-\omega_2^2x_2^2-3\omega_2x_2\cot\omega_2x_2}{3\omega_2^2 x_1}\,,
\end{equation}
where $n=n_1+n_2$.
Note that, for a boson star solution, $\omega_2x_2<\pi$ should hold.

The ratio of the Newtonian binding energy to the total mass of the boson star can
be expressed as
\begin{eqnarray}
& &\frac{E}{mn}=-\frac{S(0)\sin\omega_2x_2}{2\omega_2}\nonumber \\
& &\quad\times\left[\frac{1}{\omega_1
\sin(\omega_1x_2+\delta_1)}+\frac{(C_{11}-C_{12})^2\omega_1
\sin(\omega_1x_2+\delta_1)}{C_{11}-2C_{12}+C_{22}}
\frac{3-\omega_2^2x_2^2-3\omega_2x_2\cot\omega_2x_2}{6\omega_2^2 x_1x_2}\right]\,.
\end{eqnarray}
This expression shows the binding energy $E$ is always negative for a boson star
solution.

\begin{figure}[ht]
\centering
\includegraphics[keepaspectratio=true,width=7.5cm
]{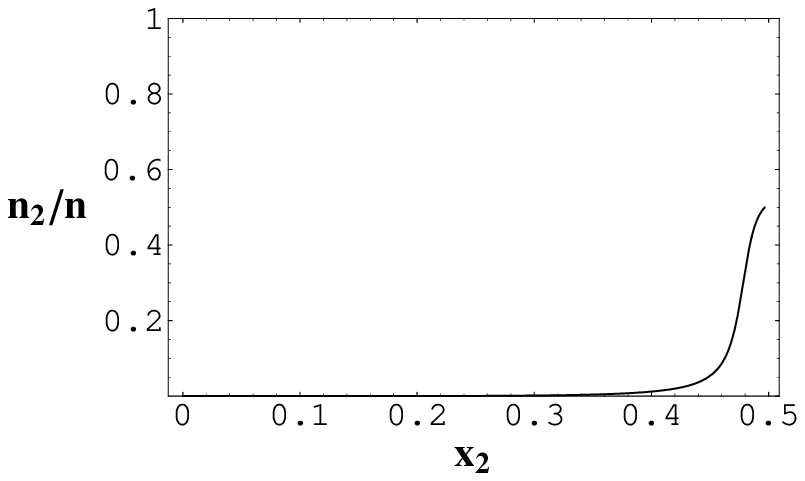}
\caption{
The fraction of the particle number of the boson 2, which lives inner region of
the boson star as a function of 
$x_2$, when $C_{11}=C_{22}=1$, $C_{12}=0.9$.
}
\label{fig03}
\centering
\includegraphics[keepaspectratio=true,width=7.5cm
]{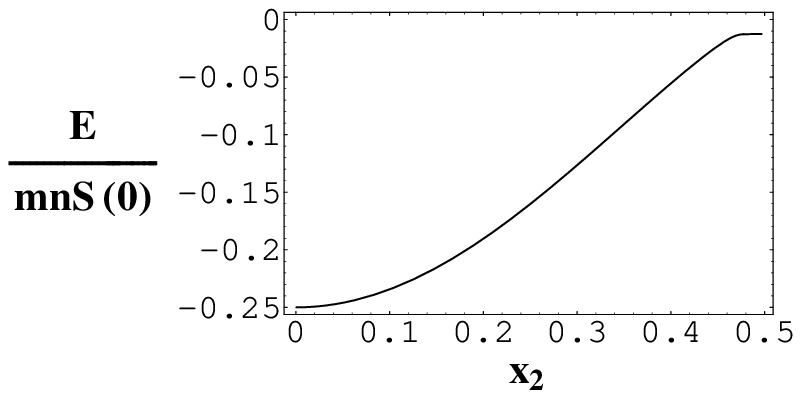}
\caption{
The ratio of the Newtonian binding energy to the total mass of the boson star as
a function of 
$x_2$, when $C_{11}=C_{22}=1$, $C_{12}=0.9$.
}
\label{fig04}
\centering
\includegraphics[keepaspectratio=true,width=7.5cm
]{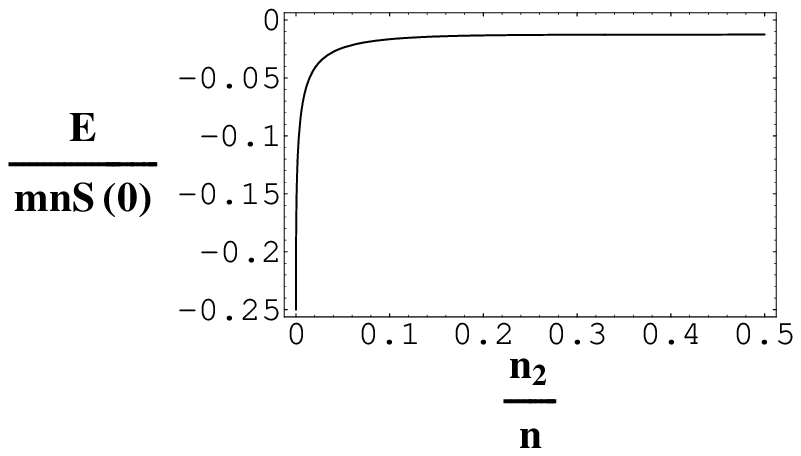}
\caption{
The ratio of the Newtonian binding energy to the
total mass of the boson star as a function of the fraction of the particle number
of the boson 2, when $C_{11}=C_{22}=1$, $C_{12}=0.9$. }
\label{fig05}
\end{figure}

We show the fraction of the particle number of the boson 2 as the function of 
$x_2$ in FIG.~\ref{fig03} and the ratio of the Newtonian binding energy to the
total mass of the boson star as
the function of 
$x_2$ in FIG.~\ref{fig04}, when $C_{11}=C_{22}=1$, $C_{12}=0.9$.
In FIG.~\ref{fig05}, we show the ratio of the Newtonian binding energy to the
total mass of the boson star as a function of the fraction of the particle number
of the boson 2, in the same case. We find that the ratio of the Newtonian binding
energy to the total mass of the boson star is nearly constant for $n_2/n>0.1$.
Note that $n_2/n$ is independent of the overall scale factor while $E/(mn)$ is
proportional to the overall scale.

Now, we consider the gravitational potential and the circular velocity. When we
vary the value of $x_2$, the shape of the boson star varies and the gravitational
potential varies at the same time. Because of the scale invariance under
(\ref{scaletransf}), we should focus our mind on the shape, not on the amount.

\begin{figure}[h]
\centering
\includegraphics[keepaspectratio=true,width=7.5cm
]{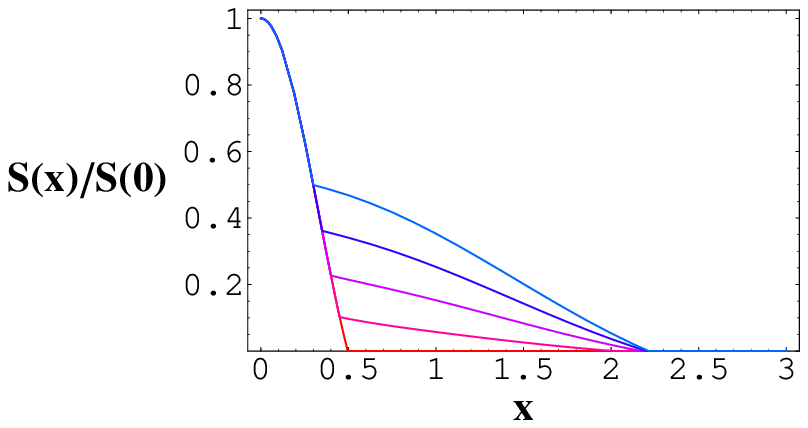}
\caption{
The behavior of the total density of scalar bosons
as the function of the rescaled distance
$x$, for $x_2=0.3,0.35,0.4,0.45,0.5$,
when $C_{11}=C_{22}=1$, $C_{12}=0.9$.
}
\label{fig06}
\centering
\includegraphics[keepaspectratio=true,width=7.5cm
]{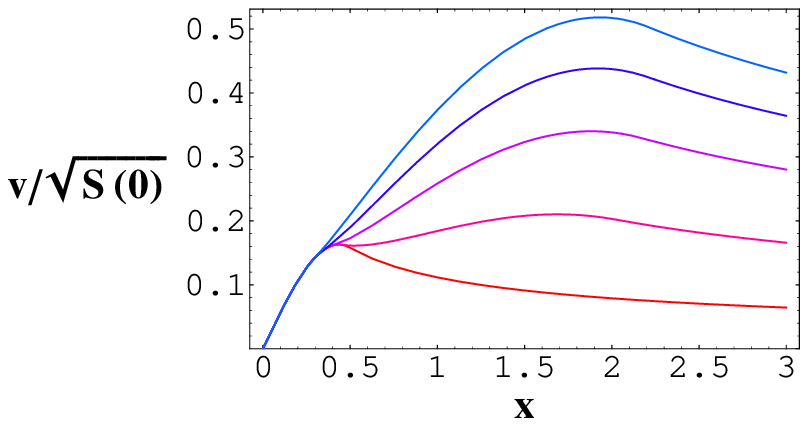}
\caption{
The galaxy rotation curves induced by the bi-scalar boson stars for
$x_2=0.3,0.35,0.4,0.45,0.5$, when $C_{11}=C_{22}=1$, $C_{12}=0.9$.
Note that overall scale can be arbitrarily chosen.
}
\label{fig07}
\end{figure}

In FIG.~\ref{fig06}, we show the various profiles of the boson stars
for $x_2=0.3,0.35,0.4,0.45,0.5$, when $C_{11}=C_{22}=1$, $C_{12}=0.9$.
The rotation speed $v$ of any object in the circular motion with radius $x$
in the Newton potential $\Phi(x)$ is proportional to $\sqrt{x \Phi'(x)}$.
In the vacuum region, $\Phi(x)$ is proportional to $-M/x$,
where
$M$ is the total mass of the boson star. Then, the rotation velocity $v$ becomes
$\propto \sqrt{M/x}$  outside the boson star. 
We exhibit $v/\sqrt{S(0)}=\sqrt{x
\Phi'(x)}/\sqrt{S(0)}$ for
$x_2=0.3,0.35,0.4,0.45,0.5$ in FIG.~\ref{fig07}. Because the potential spreads out
by the density tail due to
$\rho_1(x)$, the rotation curve can have a flat region, especially around
$x_2\approx 0.45$ in this case.
If a single scalar field model is considered,
which is realized by $\rho_2(x)\equiv 0$,
the range of the gravitational potential becomes narrow
and the rotation curve looks far from a satisfactory explanation of the
observational data.
It is interesting to point out that if the fraction of boson 2 is much larger
($\approx 0.5$, when $x_2\approx\pi\omega_2^{-1}$), the profile of the boson star
becomes much alike a single-scalar boson star.  This fact can be read from the
rotation curve in FIG.~\ref{fig07}.

So far, we have picked up an example of the couplings $C_{11}=C_{22}=1$ and
$C_{12}=0.9$. In this case, $\omega_2=6.32456$ and $\omega_1=1.41421$.
The density profile near $\rho_1\approx 0$ decreases almost linearly in
$\omega_1x$. On the other hand, the width of distribution in the core region
is determined by $\omega_2^{-1}$. Therefore, the broad gravitational potential can
be obtained if $\omega_2/\omega_1\gg 1$. In the above example, we find
$\omega_2/\omega_1=4.472$.

\begin{figure}[ht]
\centering
\includegraphics[keepaspectratio=true,width=7.5cm
]{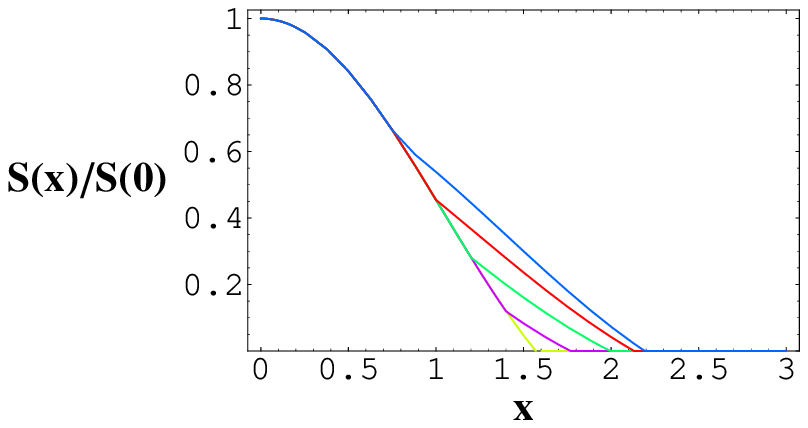}
\caption{
The behavior of the total density of scalar bosons
as the function of the rescaled distance
$x$, for $x_2=0.8,1.0,1.2,1.4,1.6$,
when $C_{11}=C_{22}=1$, $C_{12}=0$.
}
\label{fig08}
\centering
\includegraphics[keepaspectratio=true,width=7.5cm
]{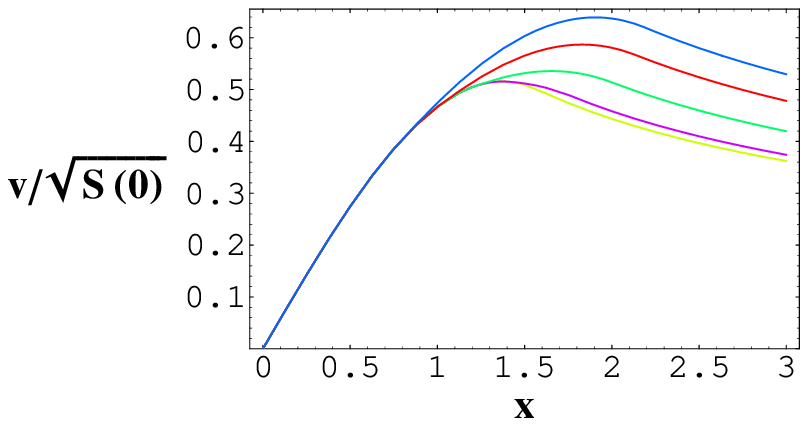}
\caption{
The galaxy rotation curves induced by the bi-scalar boson stars for
$x_2=0.8,1.0,1.2,1.4,1.6$, when $C_{11}=C_{22}=1$, $C_{12}=0$.
Note that an overall scale can be arbitrarily chosen.
}
\label{fig09}
\end{figure}

To see this necessary condition more closely, we take another example with the
couplings
$C_{11}=C_{22}=1$ and
$C_{12}=0$. In this case, we find $\omega_2=2$ and $\omega_1=1.41421$.
The density profiles and the rotation curves for various values for $x_2$ 
($x_2=0.8\sim 1.6$) are
exhibited in FIG.~\ref{fig08} and FIG.~\ref{fig09}, respectively.
From this result, we confirm the necessity of the `hierarchy' in the inverse
length scales $\omega_1$ and $\omega_2$.

In this section, we consider the simplest bi-scalar model.
We can find that the gravitational potential
is spread out by the existence of $\rho_1$, and the tail of the boson star leads
to an improvement of rotation curve of the galaxy.%
\footnote{As is known, the flat curve is realized when $S(x)\propto x^{-2}$.}
It is however difficult to obtain
the realistic galaxy rotation curve being fit to the observational data
by manipulating such a simple model.
In the next and subsequent sections, we examine multi-scalar models,
since the rotation curve may have the multi-scale structure.%
\footnote{We consider that the density of the scalar field is the total galactic
mass density, as the zero-th order approximation, in this paper.
The effect of the gravitational mass other than the boson star is briefly discussed
in Appendix \ref{appendix}.}

\section{Spherical boson stars of multi-scalar fields with a diagonal coupling
matrix}
\label{sec5}
Hereafter, we consider self-interacting $[U(1)]^N$ scalar field theory.
In this section, we examine the case that $[U(1)]^N$ symmetric scalar fields
have only self-interactions in individual fields,
and no mutual interaction to other scalar fields.
Simply speaking, we consider the case with the coupling matrix expressed as
a diagonal matrix in the present section, i.e., $C_{ij}=0$ if $i\not = j$.

To obtain the spherical static boson star solution, the equations we should solve
are now
\begin{equation}
\frac{1}{x}\frac{d^2}{dx^2}(x\Phi(x))=S(x)\,,
\label{pois}
\end{equation}
\begin{equation}
\left(\Phi(x)+
\frac{1}{2}C_{ii}\rho_i(x)\right)
{\rho}_{i}(x)=u_i{\rho}_{i}(x)\,,\quad i=1,\dots,N\,.
\end{equation}
Thus, $\rho_i$ is given by
\begin{equation}
\rho_{i}(x)=\frac{2}{C_{ii}}\left(u_i-\Phi(x)\right)>0
\qquad \mbox{or} \qquad \rho_i(x)=0\,.
\end{equation}
We take $\rho_i\neq 0$ for $i=1,\dots,k$  and
$\rho_{k+1}=\cdots=\rho_N=0$ $(1\le k\le N)$ without loss of generality. We then
define
\begin{equation}
S_k(x)\equiv\sum_{i=1}^k\rho_i(x)=\sum_{i=1}^k
\frac{2}{C_{ii}}\left(u_i-\Phi(x)\right)\,.
\end{equation}
Operating the Laplacian on the both sides of the above equation and using the
Poisson equation, we obtain
\begin{equation}
\frac{1}{x}\frac{d^2}{dx^2} (xS_k(x))=-\omega_k^2S_k(x)\,,
\quad\omega_k^2=\sum_{i=1}^k\frac{2}{C_{ii}}\,.
\label{ho}
\end{equation}
 
We assume that $\rho_i(x)$ vanishes at $x=x_i$  and the outermost surface of the
boson star is located at $x_1$, where
$0=x_{N+1}<x_N<x_{N-1}<\dots<x_2<x_1$.
All the normalized density $\rho_i$ take nonzero values in the region $0\le x\le
x_N$.
In the region $x_{k+1}\le x\le
x_k$, $\rho_N=\dots=\rho_{k+1}=0$ and then $S(x)=S_k(x)$.

The general solutions of Eq.~(\ref{ho}) are
\begin{equation}
S_k(x)=A_k\frac{\sin(\omega_k
x+\delta_k)}{x}\quad (x_{k+1}\le x\le x_k)\,,
\end{equation}
where $\delta_k$ and $A_k$ are constants. Note that $\delta_N=0$ because of
the regularity at the origin.

The gravitational potential is then given by
\begin{equation}
\Phi(x)+const.= -A_k\frac{\sin(\omega_k
x+\delta_k)}{\omega_k^2x}\quad (x_{k+1}\le x\le x_k)\,.
\end{equation}

Because of the continuity of $S(x)$ and $\Phi'(x)$, we
find the condition
\begin{equation}
\left.\frac{-x\Phi'(x)}{S(x)}\right|_{x=x_k}=\frac{\omega_kx_k\cot(\omega_k
x_k+\delta_k)-1}{\omega_k^2}=\frac{\omega_{k-1}x_k\cot(\omega_{k-1}
x_k+\delta_{k-1})-1}{\omega_{k-1}^2}\,.
\end{equation}
This is the recursive equation to determine the value of $\delta_{i-1}$
from $\delta_i$ when the other parameters are given.

First of all, we take the simplest case,
$C_{ii}=1$ and $C_{ij}=0$ for $i\not = j$, i.e., the coupling matrix
$\mathcal{C}=(C_{ij})$ is the identity matrix. In this case, because
\begin{equation}
\omega_k^2=2k\,,
\end{equation}
the relation $\omega_N>\cdots>\omega_1$ holds.
As we have seen in the previous section, we need a `hierarchy' in $\omega$
to obtain the boson star profile with a small core and a long tail.
For sufficiently large $N$, this condition is satisfied since
$\omega_N/\omega_1=\sqrt{N}\gg 1$.

\begin{figure}[ht]
\centering
\includegraphics[keepaspectratio=true,width=7.5cm
]{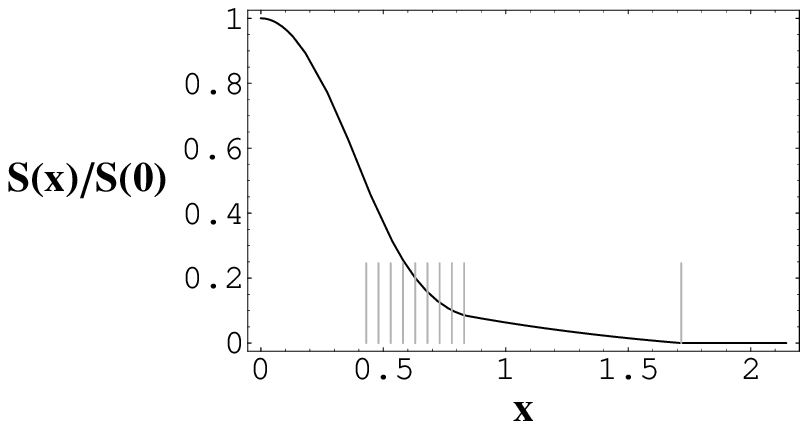}
\caption{
The behavior of the total density of ten-scalar bosons
as the function of the rescaled distance $x$ when $C_{11}=\cdots=C_{10,10}=1$,
$C_{ij}=0$. We take $\{x_{10}, x_{9}, x_8, x_7, x_6, x_5, x_4, x_3, x_2\}=\{0.43,
0.48, 0.53, 0.58, 0.63, 0.68, 0.73, 0.78, 0.83\}$ as indicated by gray vertical
lines. The
most right line indicates $x_1$, the surface of the boson star.}
\label{fig10}
\centering
\includegraphics[keepaspectratio=true,width=7.5cm
]{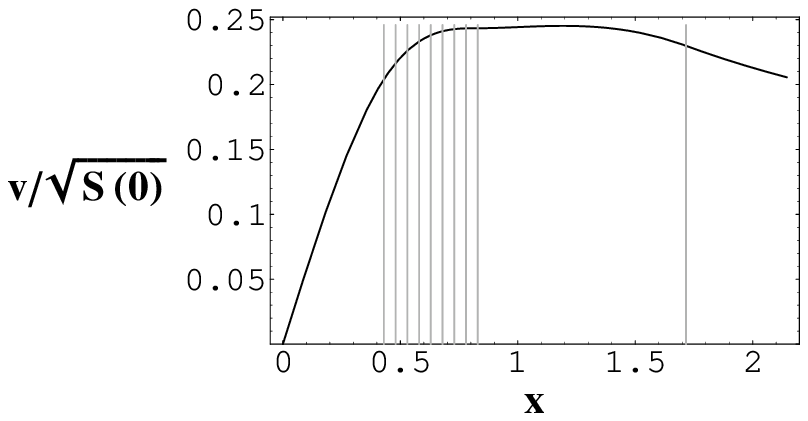}
\caption{
The galaxy rotation curves induced by the ten-scalar boson stars when
$C_{11}=\cdots=C_{10,10}=1$, $C_{ij}=0$ for $i\neq j$.
We take $\{x_{10}, x_{9}, x_8, x_7, x_6, x_5, x_4, x_3, x_2\}=\{0.43, 0.48, 0.53,
0.58, 0.63, 0.68, 0.73, 0.78, 0.83\}$ as indicated by gray vertical lines. The
most right line indicates $x_1$, the surface of the boson star. Note that overall
scale can be arbitrarily chosen. }
\label{fig11}
\end{figure}

We show a typical case of $N=10$ in FIG.~\ref{fig10} and FIG.~\ref{fig11},
where $\{x_{10}, x_{9}, x_8, x_7, x_6, x_5, x_4, x_3, x_2\}=\{0.43,
0.48, 0.53, 0.58, 0.63, 0.68, 0.73, 0.78, 0.83\}$.
In this case, $\omega_{10}/\omega_1=3.16$.

\begin{figure}[ht]
\centering
\includegraphics[keepaspectratio=true,width=7.5cm
]{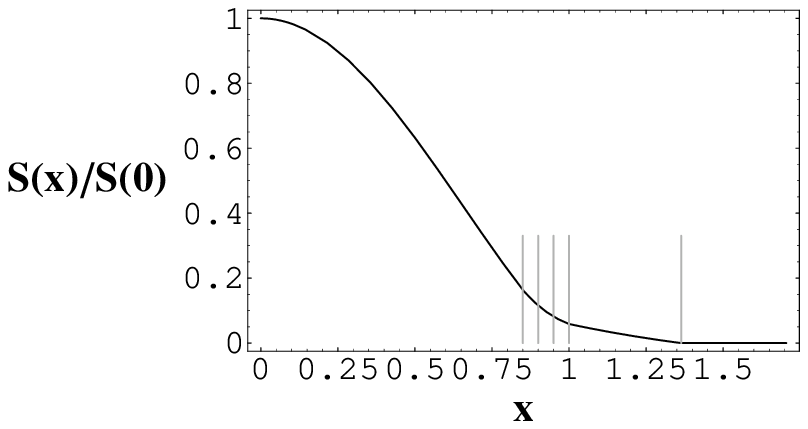}
\caption{
The behavior of the total density of five-scalar bosons
as the function of the rescaled distance $x$ when $C_{11}=\cdots=C_{55}=1$,
$C_{ij}=0$ for $i\ne j$. We take $\{x_5, x_4, x_3, x_2\}=\{0.85,
0.9, 0.95, 1\}$ as indicated by gray vertical lines. The
most right line indicates $x_1$, the surface of the boson star.}
\label{fig12}
\centering
\includegraphics[keepaspectratio=true,width=7.5cm
]{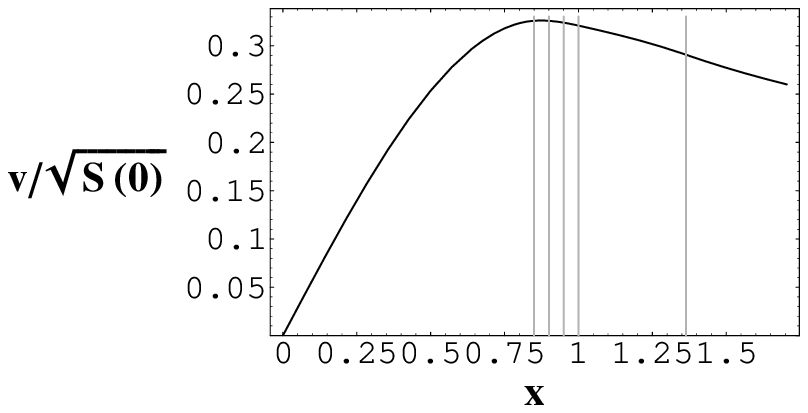}
\caption{
The galaxy rotation curves induced by the five-scalar boson stars when
$C_{11}=\cdots=C_{55}=1$, $C_{ij}=0$ for $i\neq j$.
We take $\{x_5, x_4, x_3, x_2\}=\{0.85, 0.9, 0.95,
1\}$ as indicated by gray vertical lines. The most right line indicates
$x_1$, the surface of the boson star. Note that overall scale can be arbitrarily
chosen. }
\label{fig13}
\end{figure}

For the sake of comparison, we show the case with $N=5$ in FIG.~\ref{fig12} and
FIG.~\ref{fig13}, where $\{x_5, x_4, x_3, x_2\}=\{0.85,
0.9, 0.95, 1\}$. Because $\omega_5/\omega_1$ equals $1.73$, which is not so large
enough, a small core and a long tail can hardly be obtained even if we choose a
smaller value for $x_5$, so on.

\begin{figure}[ht]
\centering
\includegraphics[keepaspectratio=true,width=7.5cm
]{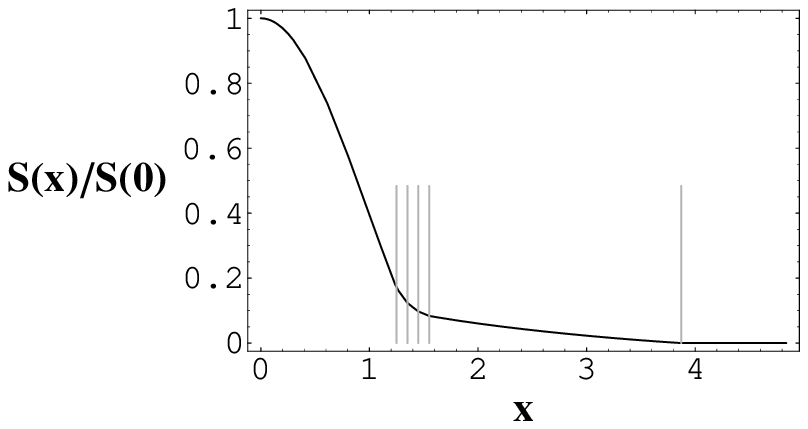}
\caption{
The behavior of the total density of five-scalar bosons
as the function of the rescaled distance $x$ when $\mathcal{C}=diag.(5, 4, 3, 2,
1)$. We take $\{x_5, x_4, x_3, x_2\}=\{1.25, 1.35, 1.45, 1.55\}$ as indicated by
gray vertical lines. The most right line indicates $x_1$, the surface of the boson
star.}
\label{fig14}
\centering
\includegraphics[keepaspectratio=true,width=7.5cm
]{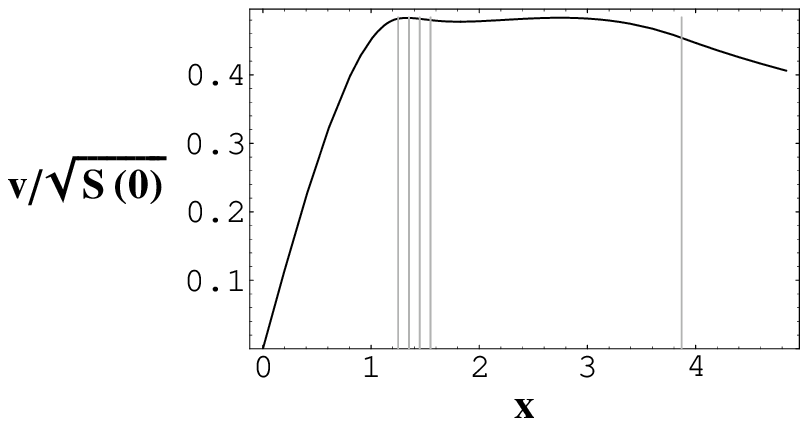}
\caption{
The galaxy rotation curves induced by the five-scalar boson stars when
$\mathcal{C}=diag.(5, 4, 3, 2, 1)$.
We take $\{x_5, x_4, x_3, x_2\}=\{1.25, 1.35, 1.45, 1.55\}$ as indicated by
gray vertical lines. The most right line indicates
$x_1$, the surface of the boson star. Note that overall scale can be arbitrarily
chosen. }
\label{fig15}
\end{figure}

Next, we consider the case that the coupling matrix is not the identity matrix but
a general diagonal matrix.
As for a typical case, the profile of a five-scalar boson star and the rotation
curves are shown  FIG.~\ref{fig14} and
FIG.~\ref{fig15}, respectively. 
Here, we take
$\mathcal{C}=diag.(5, 4, 3, 2, 1)$ and $\{x_5, x_4, x_3, x_2\}=\{1.25, 1.35, 1.45,
1.55\}$. Then we find that $\omega_5/\omega_1=3.38$.

In this section, the large number of scalar fields and/or
the large hierarchy in the couplings is necessary for a flat rotation curve,
in the case of the diagonal coupling matrix.
Both the large number of the fields and the hierarchy in couplings ruin the
simpleness of the original theory, which is implicitly desired in theoretical
models.

In the next section, we investigate more general coupling matrix including mutual
interactions among scalar fields.
Since we prefer the model with some symmetry, we propose models associated with
a certain symmetric `graph', which appears in graph theory, in the next section.

\section{a mutually-coupled scalar model associated with a graph Laplacian}
\label{sec6}
We start with a general case with non-diagonal coupling matrix
$\mathcal{C}=(C_{ij})$.
We assume that $\rho_i>0$ for $i=1,\dots, k$ and $\rho_{k+1}=\dots=\rho_{N}=0$
in the region $x_{k+1}<x<x_k$.
Then the algebraic equation for nonzero $\rho_i(x)$ can be expressed as
\begin{equation}
\Phi(x)\mathbf{w}_{[k]}+\frac{1}{2}\mathcal{C}_{[k]}\mathbf{\rho}_{[k]}(x)
=\mathbf{u}_{[k]}\,,
\end{equation}
where $\mathcal{C}_{[k]}$ is a $k\times k$ principal submatrix of the matrix
$\mathcal{C}$ defined as
\begin{equation}
\mathcal{C}_{[k]}\equiv
\left(\begin{array}{rrrr}
C_{11} & C_{12} & \cdots & C_{1k} \\
C_{21} & C_{22} & \cdots & C_{2k} \\
\vdots & \vdots & \ddots & \vdots \\
C_{k1} & C_{k2} & \cdots & C_{kk}
\end{array}\right)\,,
\end{equation}
whereas $\mathbf{w}_{[k]}$, $\mathbf{\rho}_{[k]}(x)$ and $\mathbf{u}_{[k]}$
are vectors with $k$ components given by
\begin{equation}
\mathbf{w}_{[k]}\equiv
(1,1,\dots,1)^{\rm T}\,,\quad
\mathbf{\rho}_{[k]}(x)\equiv(\rho_1(x),\rho_2(x),\dots,\rho_k(x))^{\rm T}\,,\quad
\mathbf{u}_{[k]}\equiv(u_1,u_2,\dots,u_k)^{\rm T}\,.
\end{equation}
Note that $\mathcal{C}_{[N]}=\mathcal{C}$.

In the region $x_{k+1}<x<x_k$, $\mathbf{\rho}_{[k]}(x)$ can be solved as
\begin{equation}
\mathbf{\rho}_{[k]}
=-2\mathcal{C}_{[k]}^{-1}(\Phi\mathbf{w}_{[k]}-\mathbf{u}_{[k]})\,,
\end{equation}
where $\mathcal{C}_{[k]}^{-1}$ is the inverse of $\mathcal{C}_{[k]}$.
Then, we find
\begin{equation}
S_k(x)=\sum_{i=1}^k\rho_i(x)
=-2\mathbf{w}_{[k]}^{\rm T}\mathcal{C}_{[k]}^{-1}(\Phi(x)\mathbf{w}_{[k]}-
\mathbf{u}_{[k]})\,.
\end{equation}
As in the previous sections, by using the gravitational Poisson equation,
we obtain
\begin{equation}
\frac{1}{x}\frac{d^2}{dx^2} (xS_k(x))=-\omega_k^2S_k(x)\quad
(x_{k+1}<x<x_k)\,,
\end{equation}
where
\begin{equation}
\omega_k^2=2\mathbf{w}_{[k]}^{\rm T}\mathcal{C}_{[k]}^{-1}\mathbf{w}_{[k]}\,.
\end{equation}
Now, since we obtain $\omega_k^2$, we can evaluate the profile of the boson star
and the gravitational potential by the same method as in the previous sections.


In the rest of this section, we consider a concrete model whose coupling matrix
is represented by a graph Laplacian, which appears in texts of spectral graph
theory
\cite{mohar1,mohar2,mohar3,merris,GR,CRS}. 

Let $G(\mathcal{V},\mathcal{E})$ be a graph with a vertex set $\mathcal{V}$ and
an edge set $\mathcal{E}$. The set of edges connects the vertices.
A pair of vertices $v$ and $v'$ are said to be adjacent, denoted $v\sim v'$,
if there exists an edge which connects $v$ and $v'$.
The degree of a vertex $v$, denoted $deg(v)$, is the number of edges directly
connected to $v$.

The graph Laplacian $\triangle(G)$ of the graph $G$ is defined by 
\begin{equation}
(\triangle(G))_{vv'}=
\left\{
\begin{array}{cl}
deg(v) & \mbox{if  }~ v=v' \\
-1 & \mbox{if  }~ v\sim v'  \\
0 & \mbox{otherwise}
\end{array}\right.\,,
\end{equation}
where $v, v'\in \mathcal{V}$.

Now, we take a model whose coupling matrix can be written as
\begin{equation}
\mathcal{C}=\mathcal{I}+\gamma \triangle(G)\,,
\label{cm}
\end{equation}
where $\mathcal{I}$ is the $N\times N$ identity matrix
and $\gamma$ is a non-negative constant.
Here, we suppose that the graph $G$ has $N$ vertices.
We inversely find the scalar quartic interaction in this model has the form
\begin{equation}
V(|\phi_i|^2)\propto\sum_{v\in{\cal V}}|\phi_v|^4+\gamma
\sum_{v\sim v'}(|\phi_v|^2-|\phi_{v'}|^2)^2\,,
\end{equation}
where the second sum is done once over all adjacent pairs, and then we find the
potential is obviously positive-semidefinite. The use of the graph Laplacian
guarantees the positivity of the potential energy in the model.

\begin{figure}[ht]
\centering
\includegraphics[height=5cm
]
{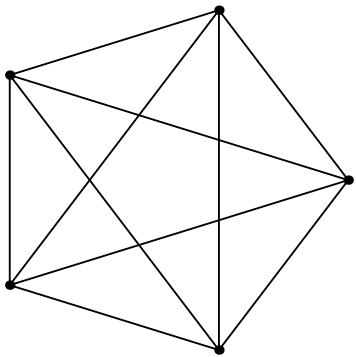}
\caption{
A complete graph $K_5$.
}
\label{fig16}
\end{figure}

First of all, we adopt a complete graph $K_N$ as a maximally symmetric graph.
In a complete graph, each vertex is adjacent to all the other vertices
(FIG.~\ref{fig16}).
The model based on a complete graph is most democratic, because the potential is
invariant under any exchange of scalar field species, in other words, vertices in
a complete graph have a symmetry under the symmetric group $S_N$.
The graph Laplacian of the complete graph with $N$ vertices is
\begin{equation}
\triangle(K_{N})=
\left(\begin{array}{ccccc}
N-1 & -1 & -1 & \cdots & -1 \\
-1 & N-1 & -1 & \cdots & -1 \\
-1 & -1 & N-1 & \cdots & -1\\
\vdots & \vdots & \vdots & \ddots & \vdots \\
-1 & -1 & -1 & \cdots & N-1
\end{array}\right)\,.
\end{equation}

To obtain $\omega_k^2$, we have to calculate the matrix $\mathcal{C}_{[k]}^{-1}$.
We consider the eigenvectors of the matrix $\mathcal{C}_{[k]}$,
which is denoted by $\mathbf{v}_{[k]}^a$ $(a=0,\dots,k-1)$. They satisfy
\begin{equation}
\mathcal{C}_{[k]}\mathbf{v}_{[k]}^a=\lambda_{[k]}^a\mathbf{v}_{[k]}^a\,,\quad
(\mathbf{v}_{[k]}^a)^{\rm T}\mathbf{v}_{[k]}^b=\delta^{ab}\,,
\end{equation}
where $\lambda_{[k]}^a$ is the eigenvalues of the matrix $\mathcal{C}_{[k]}$.
Then, we can express the inverse matrix using the eigenvalues and the eigenvectors
of the matrix as
\begin{equation}
\mathcal{C}_{[k]}^{-1}=\sum_{a}\mathbf{v}_{[k]}^a\frac{1}{\lambda_{[k]}^a}
(\mathbf{v}_{[k]}^a)^{\rm T}\,.
\end{equation}
In the present case that $\mathcal{C}=\mathcal{I}+\gamma \triangle(K_N)$,
the vector $\frac{1}{\sqrt{k}}\mathbf{w}_{[k]}$ is an eigenvector of 
$\mathcal{C}_{[k]}$ for any $k$, and the associated eigenvalue is $1+\gamma(N-k)$.
Therefore,
\begin{equation}
\omega_k^2=2\mathbf{w}_{[k]}^{\rm T}\mathcal{C}_{[k]}^{-1}\mathbf{w}_{[k]}
=2\mathbf{w}_{[k]}^{\rm T}\mathbf{w}_{[k]}\frac{1}{k[1+\gamma(N-k)]}
\mathbf{w}_{[k]}^{\rm T}\mathbf{w}_{[k]}
=\frac{2k}{1+\gamma(N-k)}\,,
\end{equation}
where we used the orthogonal relation among the eigenvectors.

The necessary condition for a `small core and long tail' boson star solution
is that the ratio $\omega_N/\omega_1$ is much larger than unity, actually
$\omega_N/\omega_1\stackrel{>}{\sim} 3\sim 4$.
In the present model, we find
\begin{equation}
\frac{\omega_N}{\omega_1}=\sqrt{N(1+\gamma(N-1))}\sim \gamma^{1/2}N\,,
\end{equation}
so, the choice of $\gamma\simeq 1$ and $N\simeq 3\sim 4$ satisfies the condition.

\begin{figure}[ht]
\centering
\includegraphics[keepaspectratio=true,width=7.5cm
]{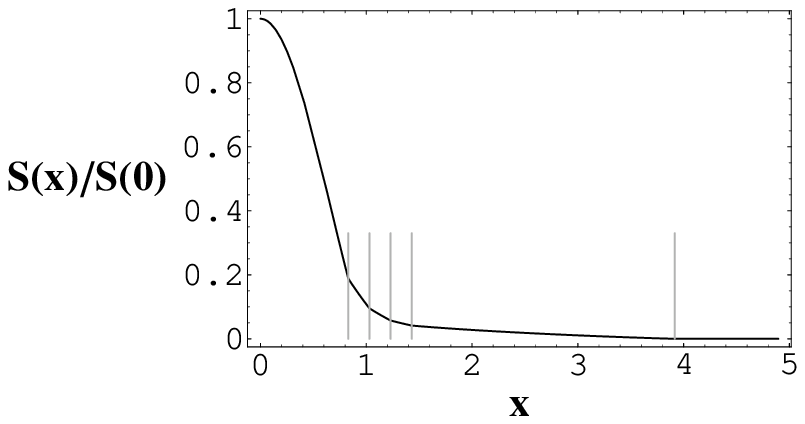}
\caption{
The behavior of the total density of five-scalar bosons
as the function of the rescaled distance $x$ in the model associated with a
complete graph $K_5$ and $\gamma=1$. We take $\{x_5, x_4, x_3, x_2\}=\{0.83, 1.03,
1.23, 1.43\}$ as indicated by gray vertical lines. The most right line indicates
$x_1$, the surface of the boson star.}
\label{fig17}
\centering
\includegraphics[keepaspectratio=true,width=7.5cm
]{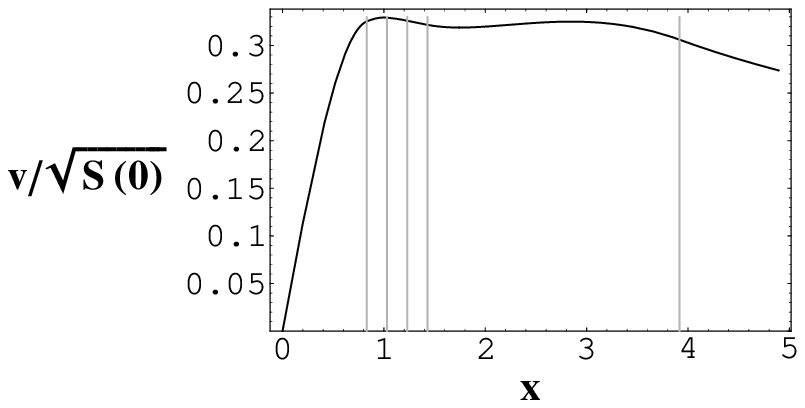}
\caption{
The galaxy rotation curves induced by the five-scalar boson stars 
in the model associated with a complete graph $K_5$ and $\gamma=1$.
We take $\{x_5, x_4, x_3, x_2\}=\{0.83, 1.03, 1.23, 1.43\}$ as indicated by
gray vertical lines. The most right line indicates
$x_1$, the surface of the boson star. Note that overall scale can be arbitrarily
chosen. }
\label{fig18}
\end{figure}

\begin{figure}[ht]
\centering
\includegraphics[keepaspectratio=true,width=7.5cm
]{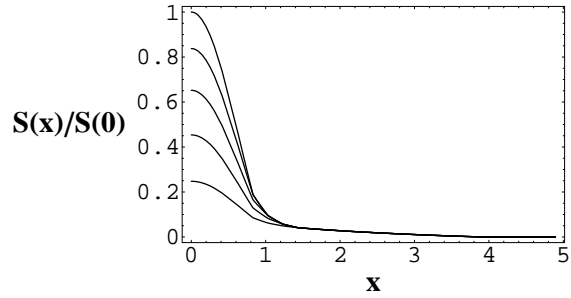}
\caption{
The contribution of each boson species
 in the model associated with $K_5$
and $\gamma=1$. We take $\{x_5, x_4, x_3, x_2\}=\{0.83, 1.03, 1.23, 1.43\}$.
The curves indicate $S_k(x)=\sum_{i=1}^k\rho_i(x)$, $k=5$ to $1$, from the upper to
the lower. }
\label{fig19}
\end{figure}

We demonstrate the calculation for $N=5$, $\gamma=1$, 
and $\{x_5, x_4, x_3, x_2\}=\{0.83, 1.03, 1.23, 1.43\}$, and show the results
in FIG.~\ref{fig17} and FIG.~\ref{fig18}.
By integrating densities, we can evaluate the particle number of each scalar boson.
The fraction of species are
$\{n_1/n,n_2/n,n_3/n,n_4/n,n_5/n\}=\{0.786, 0.0788, 0.0620, 0.0453,
0.0283\}$, where $n=\sum_{i=1}^5 n_i$.
Although the value of $n_1/n$ is larger than those of others, because of the long
tail distribution of the boson star, this composition is not so unnatural.
The contribution of each boson species are shown in FIG.~\ref{fig19}.

Thus, we have found that the model with the coupling matrix associated with the
graph Laplacian of the complete graph leads to flat rotation curves in a natural
setting, for comparatively small number of scalar fields which have 
symmetry under any exchange of species.
Moreover, a small but finite number of fields brings about a diversity of
galaxy rotation curves, as observations indicate. It can be recognized
that the variation on diverse galaxies is due to the set of fractions of the
particle numbers.


\begin{figure}[ht]
\centering
\includegraphics[height=5cm
]
{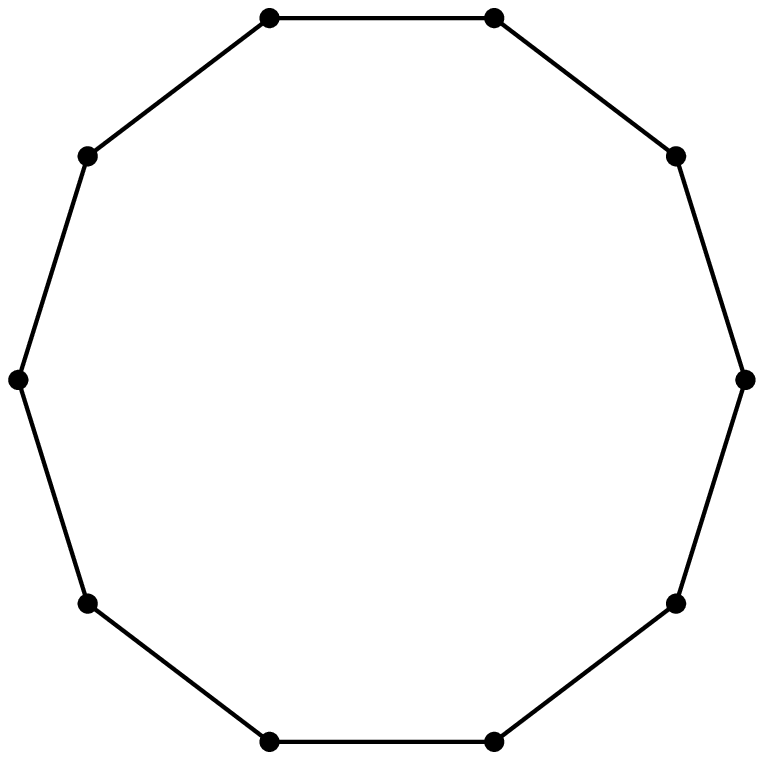}
\caption{
A cycle graph $C_{10}$.
}
\label{fig20}
\hspace{5mm}
\centering
\includegraphics[height=5cm
]
{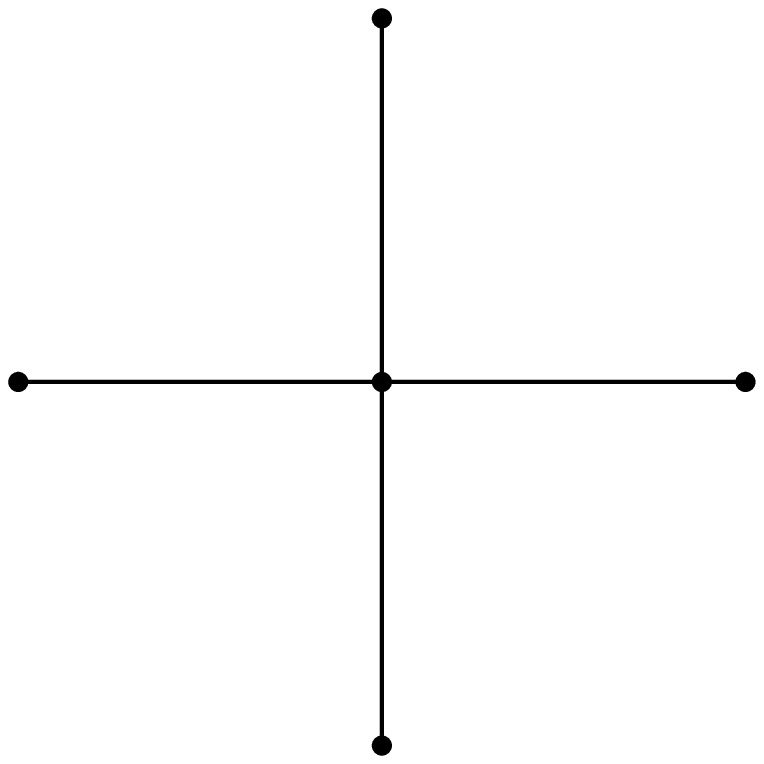}
\caption{
A star graph with five vertices.
}
\label{fig21}
\end{figure}

Next, we consider the use of the other graphs in the coupling matrix.
We now consider a cycle graph $C_N$ (FIG.~\ref{fig20}).
The coupling matrix is given in the same form Eq.~(\ref{cm}), as previously
considered. The model based on a cycle graph still has discrete symmetry under
$\phi_i\rightarrow \phi_{i+l}~ (\mbox{mod } N)$, where $l$ is an integer.

The graph Laplacian of a cycle graph $C_N$ takes the form
\begin{eqnarray}
\triangle(C_{N})=
\left(\begin{array}{cccccc}
2 & -1 & 0 & \cdots & 0 &-1 \\
-1 & 2 & -1 & \cdots & 0 &0\\
0 & -1 & 2 & \cdots & 0 & 0\\
\vdots & \vdots & \vdots & \ddots & \vdots &\vdots\\
0 & 0 & 0 & \cdots & 2 & -1\\
-1 & 0 & 0 & \cdots & -1 & 2 
\end{array}\right)\,.
\end{eqnarray}

In the model associated with $\triangle(C_N)$, we find
\begin{equation}
\omega_N^2=2\mathbf{w}_{[N]}^{\rm T}\mathcal{C}^{-1}\mathbf{w}_{[N]}=2N\quad
\mbox{and}\quad
\omega_1^2=2{C}_{11}^{-1}=\frac{2}{1+2\gamma}\,.
\end{equation}
Note that $\frac{1}{\sqrt{N}}\mathbf{w}_{[N]}$ is the eigenvector belonging to
zero eigenvalue for any simple graph Laplacian.
Therefore, we find%
\footnote{If $deg(v)=d$ for all $v\in\mathcal{V}$, the graph is called
a $d$-regular graph. The cycle graph is a $2$-regular graph.
For the model associated with $d$-regular graph, one can find
$\omega_N/\omega_1=\sqrt{N(1+\gamma d)}$.}
\begin{equation}
\frac{\omega_N}{\omega_1}=\sqrt{N(1+2\gamma)}\sim\sqrt{N}\,.
\end{equation}
In general, the necessary condition in this model is more severe than that in the
model associated with a complete graph, but a sufficiently large number of scalar
fields satisfies the condition.
In a loosely connected graph such as a cycle graph,
each degree of the vertex is much smaller than $N-1$. Thus, $\omega_1$ tends to
be larger than the case with a complete graph.
In the star graph with $N$ vertices (FIG.~\ref{fig21}), the maximal degree of the
vertex is
$N-1$, which is the same as that in a complete graph.
Then, in the model based on the star graph,
the ratio ${\omega_N}/{\omega_1}$ takes the
same value as that in the model of the complete graph $K_N$ with the same number
$N$ of scalar species. However, the model associated with the star graph with $N$
vertices has less symmetry, i.e. symmetric under the symmetric group $S_{N-1}$
rather than
$S_N$.

Recently, models with mass hirarchy generated by the `clockwork mechanism'
\cite{CI,KaRa,GM1,FPRT,KeRi,HTT,teresi,AD,CGS,GM2} are eagerly investigated by many
authors. If it is feasible to use the same structure in our coupling matrix
instead of their mass matrices, we would have an interesting model for boson stars
constructed by several fields.

\section{Summary and outlook}
\label{so}
In the present paper, we have examined the Newtonian boson star with many $U(1)$
charges  in the large coupling limit.
The explanation of 
rotation curves of galaxies by gigantic boson stars is improved in this model.

The necessary condition for a flat rotation curve is that $\omega_N$ is several
times larger than $\omega_1$ in terms of our model parameter.
This is naturally led from the model with a large number of scalar field
and/or the model whose coupling matrix is associated with the graph Laplacian of
the complete graph. The latter model has larger symmetry on several
scalar fields. It is worth pointing out that several species of bosons are
necessary to make a variety in the rotation curves of different galaxies,
even including dwarf galaxies.%
\footnote{We may need, however, a large number $N$ naturally to obtain hierarchical
scales.}

In the present paper, we have concentrated ourselves on the galactic boson stars.
The structure of small multi-scalar boson stars is still similar because of the
scale invariance of our models. It will be interesting to find some new aspects
of the boson stars in various circumstances, such as in a collision of
multi-scalar boson stars.%
\footnote{The collision of boson stars has been studied in Refs.~\cite{BG,GG}, for
example.}

Although the model we have demonstrated is rather a toy model and has only been
focused as in the large coupling limit, it is the simplest effective theory of
multi-scalar boson stars.
As a variant of the model, it is interesting to consider
the case that there exist other fields which do not participate the ingredient
of the boson star but affect on the creation or decay of boson stars.
Anyway, we would like to find the relevance of the
multi-scalar theory to particle physics, and wish to clarify its role or
significance.

In future work, we will consider
general relativistic boson stars and graph-oriented models with
many charges or arbitrary couplings. 
We also have much interest on boson stars in plausible models in which finite or
zero couplings.
It is known that single-scalar boson stars with finite self-couplings can also be
approximated analytically by connecting the exponential tail in the asymptotic
region of the boson star \cite{GA}.
Since the rotation curve is most sensitive on the tail of the boson star,
analyses of models with finite self-couplings should be performed as in the next
step. Time dependent solutions or oscillations of multiple scalar fields are also
interesting. We wish to investigate these subjects elsewhere. 


\appendix

\section{a multi-scalar boson star with an external gravitational source}
\label{appendix}
The realistic galaxies have bulges and halos of stars and gas, which are
gravitational sources and are expected not to interact with scalar fields in
our model (and other unknown dark stuff).

In the presence of the nonnegligible external gravitational source,%
\footnote{We neglect further back reaction to the source from the gravitation of
scalar bosons.}
 whose
energy density is given by $\rho_{ext}$ other than the scalar fields,
we add an inhomogenious term in the right hand side of Eq.~(\ref{pois}) as
\begin{equation}
\frac{1}{x}\frac{d^2}{dx^2}(x\Phi(x))=S(x)+\bar{\rho}_{ext}(x)\,,
\end{equation}
where $\bar{\rho}_{ext}=4\pi G\Lambda/m^2 \rho_{ext}$.
Although the normalization seems to be strange, the ratio of energy densities
are nevertheless simply given by $\bar{\rho}_{ext}/S$.
Now, we should solve the differential equations
\begin{equation}
\frac{1}{x}\frac{d^2}{dx^2}(xS_k(x))+\omega_k^2S_k(x)
=-\omega_k^2\bar{\rho}_{ext}(x)
\quad (x_{k+1}<x<x_k)\,,
\end{equation}
and
the special solution for the inhomogeneous term is found to be
\begin{equation}
-\frac{\omega_k}{x}\int_{x_{k+1}}^x t\bar{\rho}_{ext}(t)\sin\omega_k(x-t)dt
\quad (x_{k+1}<x<x_k)\,.
\end{equation}

Since a main subject of the present paper is an interest in analytical
study, we consider a simple example in this appendix.
We now assume
\begin{equation}
\bar{\rho}_{ext}(x)=A_e\frac{\sin\omega_e x}{x}\quad (0<x<\pi/\omega_e)\,,
\end{equation}
where $A_e$ and $\omega_e$ are constants and $\bar{\rho}_{ext}(x)=0$ for
$x>\pi/\omega_e\equiv x_{N+1}$. We further assume $\omega_e>\omega_N$ for
simplicity. 
Then, we find
\begin{equation}
S_{N+1}(x)=A_N\frac{\sin\omega_N
x}{x}+\frac{A_e}{\omega_e^2/\omega_N^2-1}\frac{\sin\omega_e x}{x}\quad
(0<x<\pi/\omega_e=x_{N+1})\,,
\end{equation}
where $A_N$ is an arbitrary constant.
We have only to find the total solutions by the connection condition of $-x\Phi'/S$
as in Sec.~\ref{sec5}.

\begin{figure}[ht]
\centering
\includegraphics[height=5cm
]
{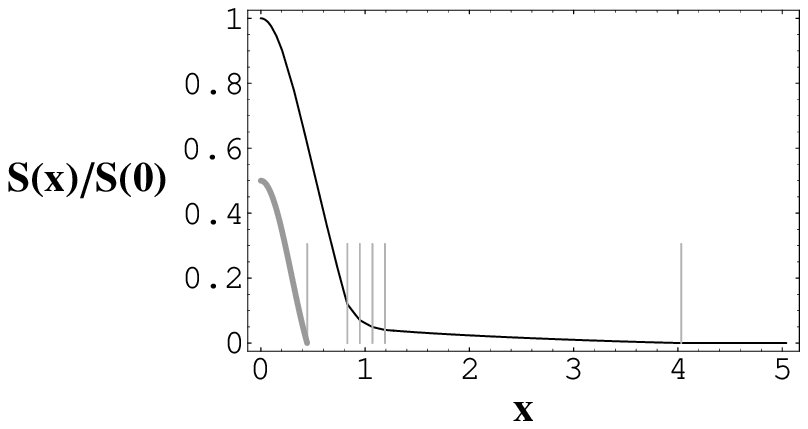}
\caption{
The behavior of the total density of five-scalar bosons
as the function of the rescaled distance $x$ in the model associated with a
complete graph $K_5$ and $\gamma=1$, in the presence of the external gravitational
source (see text) with $\omega_e=\sqrt{5}\omega_5$ and
$\bar{\rho}_{ext}(0)/S(0)=0.5$. The gray curve in the figure shows
$\bar{\rho}_{ext}(x)$. We take
$\{x_6,x_5, x_4, x_3, x_2\}=\{0.44, 0.83, 0.95, 1.07, 1.19, 1.31\}$ as indicated
by gray vertical lines. The most right line indicates
$x_1$, the surface of the boson star.}
\label{figA1}
\hspace{5mm}
\centering
\includegraphics[height=5cm
]
{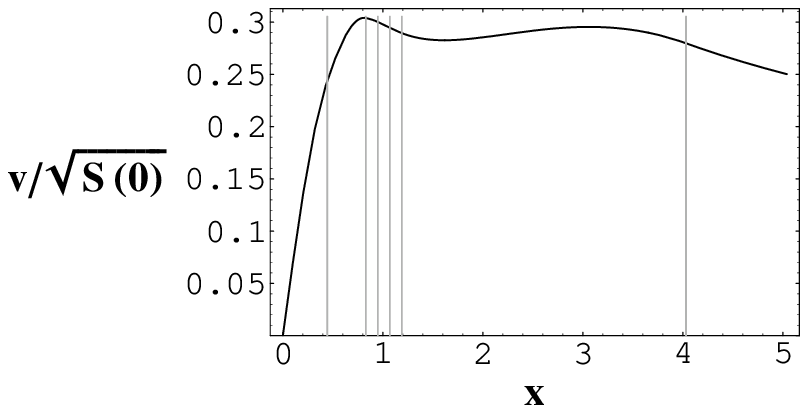}
\caption{
The galaxy rotation curves induced by the five-scalar boson stars 
in the model associated with a complete graph $K_5$ and $\gamma=1$, in the presence of the external gravitational
source (see text) with $\omega_e=\sqrt{5}\omega_5$ and
$\bar{\rho}_{ext}(0)/S(0)=0.5$. We take $\{x_5, x_4, x_3, x_2\}=\{0.44, 0.83,
0.95, 1.07, 1.19, 1.31\}$ as indicated by gray vertical lines. The most right line
indicates
$x_1$, the surface of the boson star. Note that overall scale can be arbitrarily
chosen. }
\label{figA2}
\end{figure}

A result in the model associated with $K_5$ in Sec.~\ref{sec6} is shown in
FIG.~\ref{figA1} and FIG.~\ref{figA2}, where we set
$\omega_e=\sqrt{5}\omega_5$,
$\bar{\rho}_{ext}(0)/S(0)=0.5$ and $\{x_5, x_4, x_3, x_2\}=\{0.44, 0.83, 0.95,
1.07, 1.19, 1.31\}$.
The most appropriate parameter set for a flat rotation curve, of course, changes
slightly by the effect of the gravitational source.
Conversely, it would be said that the density profle from the present model can be
adjusted to various distributions of ordinary matter in many galaxies.


\bibliographystyle{apsrev4-1}

\end{document}